\documentclass[aps,preprint]{revtex4}%
\usepackage{amsfonts}
\usepackage{amsmath}
\usepackage{amssymb}
\usepackage{graphicx}%
\providecommand{\U}[1]{\protect\rule{.1in}{.1in}}
\newtheorem{theorem}{Theorem}

\newenvironment{proof}[1][Proof]{\noindent\textbf{#1.} }{\ \rule{0.5em}{0.5em}}
\ifx\pdfoutput\relax\let\pdfoutput=\undefined\fi
\newcount\msipdfoutput
\ifx\pdfoutput\undefined\else
\ifcase\pdfoutput\else
\ifx\paperwidth\undefined\else
\ifdim\paperheight=0pt\relax\else\pdfpageheight\paperheight\fi
\ifdim\paperwidth=0pt\relax\else\pdfpagewidth\paperwidth\fi
\fi\fi\fi
\begin{document}
\preprint{Internal Communication }
\title{Universes with and without a center}
\author{Yukio Tomozawa}
\affiliation{Michigan Center for Theoretical Physics}
\affiliation{Randall Laboratory of Physics}
\affiliation{University of Michigan}
\affiliation{Ann Arbor, MI. 48109-1040, USA}
\date{\today }

\begin{abstract}
Two types of universe, with and without a center, are discussed; and their
implications for the observed cmb (cosmic microwave background radiation)
dipole are described.\ Theorems useful for understanding the cause for a cmb
dipole are presented. Using the theorems stated and all existing data, a
computation for the location of the center of the universe has been performed.
In the universe without center, however, a complication arises by the
appearance of the apparent center after the expansion of the universe has proceeded.

\end{abstract}

\maketitle


\section{\label{sec:level1}Introduction}

There are two types of cosmology in the realm of the Friedman universe, one
with a center for expansion and the other without a center. Hubble's law,
\textbf{v }= H$_{\text{0}}$ \textbf{r}, yields the relationship,
\textbf{v}$_{\text{2}}$ - \textbf{v}$_{\text{1}}$= H$_{\text{0}}$ (
\textbf{r}$_{\text{2}}$ - \textbf{r}$_{\text{1}}$) for any two galaxies with
positions and velocities, \textbf{r}$_{\text{1}}$, \textbf{v}$_{\text{1}}$ and
\textbf{r}$_{\text{2}}$ , \textbf{v}$_{\text{2}}$ respectively, where
H$_{\text{0}}$ = 100 h km/s-Mpc is the Hubble constant (with h = 0.5
\symbol{126}0.85). For convenience of discussion, we assume the value of
H$_{\text{0}}$ to be 70.0 km/s-Mpc in this article. This equation implies that
every point appears to be the center of the expansion. In other words, both
types of cosmology yield the same conclusion, as far as the Hubble law is
concerned. However, the observed cmb dipole has different implications for the
two types of cosmology and observational differences will be presented.

\section{Two Types of Cosmology}

In the Friedman universe,%
\begin{equation}
ds^{2}=dt^{2}-a(t)^{2}(dr^{2}/(1-kr^{2})+r^{2}d\theta^{2}+r^{2}\sin^{2}%
(\theta)d\phi^{2}),
\end{equation}
with an appropriate source, $T_{\nu}^{\mu}$ , there are two types of
interpretation for the radial coordinate, $r$.

I) B-type Universe

The universe resides on the surface of an expanding balloon (B for balloon).
The center does not exist in the universe. (It exists outside the universe.)
The coordinate origin can be chosen at any point but there is no special
significance for such a choice. The Hubble law is naturally built into the
framework. There are no velocities associated with individual points of the
universe, but the relative distance and relative velocity of any two points
increase with the expansion of the balloon. Apparently, many physicists
subscribe to this type of universe as their image.

II) C-type Universe

The origin of the radial coordinate has a physical meaning as the point where
the expansion started. Each point of the universe has a Hubble flow velocity
relative to the origin that is proportional to the distance from the origin (C
for center). As discussed in the introduction, the linearity of the Hubble law
makes every point in the universe look like a center for the expansion of the
universe, even though the center of the universe exists.

I will discuss the implications of the observed cmb dipole for cosmologies of
B-type and C-type. An important point is how to discriminate the two types of
cosmology observationally.

\section{The cmb Dipole in a B-type Universe}

Each point is equivalent relative to a distant cmb emittor and there is no cmb
dipole at any point except possibly that due to a peculiar velocity in a
cluster. Let me state this in the form of a theorem.

\begin{theorem}
No point in a B-type universe observes a cmb dipole.

\begin{proof}
This is obvious because of symmetry between opposite directions.
\end{proof}
\end{theorem}

However, the next theorem might be a surprise to some.

\begin{theorem}
A peculiar velocity at a point in a B-type universe does not produce a cmb dipole.

\begin{proof}
Let an object at point A have a peculiar velocity \textbf{v}. There is a point
A$_{\mathbf{v}}$ in the direction of \textbf{v} which has a velocity
\textbf{v} relative to A. Peculiar velocities at points A and A$_{\mathbf{v}}$
have the same motion, and point A$_{\mathbf{v}}$ has no cmb dipole by Theorem
1, so a peculiar velocity at A does not observe a cmb dipole. In order to help
visualize this, let us imagine a structure extended from A$_{\mathbf{v}}$ to
A. Since a structure does not expand along with the universe, all points on it
should observe the same cmb dipole, a vanishing cmb dipole in this case. A
peculiar velocity at point A and at a point on the extended structure
coinciding with A have the same velocity and the same location, so they should
have the same cmb dipole. Since the point on the extended structure has zero
cmb dipole, the theorem has been proved.
\end{proof}
\end{theorem}

The content of this section is true in the early stage of the universe
expansion. At the later stage of the universe expansion, however, there is a
possibility of the appearance of an apparent center for a B-type universe. It
will be discussed at the end of the article.

\section{The CMB Dipole in a C-type Universe}

In this universe, each point is moving away from a fixed point within it, the
center, with velocity \textbf{v}, which may be called the Hubble flow
velocity. The velocity is written in terms of the distance \textbf{r} of the
point from the center as%
\begin{equation}
\mathbf{v}=H_{0}\mathbf{r,}%
\end{equation}
where H$_{0}$ is the Hubble constant.

\begin{theorem}
A point in a C-type Universe observes a cmb dipole \textbf{v}, where
\textbf{v} is the Hubble flow velocity.

\begin{proof}
Let us denote the velocity of a cmb emittor in the direction of the Hubble
flow velocity by \textbf{V}. From the size of the present universe, V is close
to speed of light, c. The relative velocity of the cmb emittor and an observer
with Hubble flow velocity \textbf{v }is given by%
\begin{equation}
(V-v)/(1-\frac{Vv}{c^{2}})=V-v+\frac{(V-v)V}{c^{2}}v=V-v+v \label{hf1}%
\end{equation}
to linear approximation in v. Here, we used
\begin{equation}
V^{2}/c^{2}\approx1
\end{equation}
A cmb emittor in the direction opposite to the Hubble flow should have
velocity $V-2v$ in order to reach the observer at the same time as the one
from the opposite direction. The relative velocity in this direction is
\begin{equation}
(V-2v+v)/(1+\frac{(V-2v)v}{c^{2}})=V-v-\frac{(V-v)(V-2v)}{c^{2}}v=V-v-v
\label{hf2}%
\end{equation}
The difference between Eq (\ref{hf1}) and Eq. (\ref{hf2}) shows a cmb dipole
\textbf{v}, which is identical to the Hubble flow velocity.
\end{proof}
\end{theorem}

The following shows the effect of a peculiar velocity in a C-type Universe.

\begin{theorem}
A peculiar velocity in a C-type universe yields a cmb dipole with velocity
equal to the vector sum of the Hubble flow velocity and the peculiar velocity.

\begin{proof}
The proof is very similar to that of Theorem 2 in the previous section. The
only difference is the nature of the rest of the Universe concerning the cmb
dipole. The motion of a peculiar velocity, \textbf{v}$_{p}$, at a Hubble flow
velocity, \textbf{v, }is identical to that at a Hubble flow velocity,
\textbf{v} + \textbf{v}$_{p}$. Since the latter has a cmb dipole, \textbf{v} +
\textbf{v}$_{p}$, a peculiar velocity contributes to a cmb dipole,
\textbf{v}$_{p}$. If necessary, consider an extended structure from the
location of the Hubble flow velocity, \textbf{v} + \textbf{v}$_{p}$, to the
location of a Hubble flow velocity, \textbf{v}.
\end{proof}
\end{theorem}

\section{Temporary conclusion}

Summarizing the theorems in the previous sections, I can state the following.

1) In a B-type universe, nobody observes a cmb dipole. Not even a peculiar
velocity yields a cmb dipole.

2) In a C-type universe, every body observes a cmb dipole which is equal to
its Hubble flow velocity. Away from the center, the magnitude of the cmb
dipole increases with distance from the center. With a peculiar velocity
presenting, the cmb dipole is the vector sum of the Hubble flow velocity and
the peculiar velocity. At the center, the cmb dipole is zero and only one
point in the universe has such a property. If a cmb dipole and a peculiar
velocity coincide with each other, one has to reside at the center.

Now that a cmb dipole has been observed\cite{dipole1},\cite{dipole2}, one has
to conclude that there should exist a center of the universe.

I) If the observed cmb dipole and the peculiar velocity of the solar system
coincide, as is assumed among some physicists, the solar system must reside at
the center of the universe. As explained elsewhere\cite{center}, this is not
the case. The solar system moves towards the Virgo cluster\cite{sciama1}%
,\cite{sciama2},\cite{sciama3} and the Virgo cluster moves towards the Great
Attractor\cite{ga} (hereafter called GA)

II) A B-type universe is not compatible with the observation. A C-type
universe must be accepted.

III) The observed cmb dipole is the vector sum of the Hubble flow and peculiar
velocities. It is a measure of the effective distance from the center of the
universe. It is remarkable that the measurement of the cmb dipole is nothing
but the measurement of this distance.

IV) The magnitudes of the observed cmb dipole and the above mentioned peculiar
velocities are relatively small. This means that we are relatively close to
the center of the universe.

This conclusion, however, will be altered later in the article.

\section{The position of the center}

The observed cmb dipole for blue shift is expressed as\cite{dipole1}%
,\cite{dipole2}%
\begin{equation}
v(dipole)=371\pm0.5\text{ }km/s,\text{ \ \ \ }l=264.4\pm0.3%
{{}^\circ}%
,\text{ \ \ \ }b=48.4\pm0.5%
{{}^\circ}%
, \label{dipole}%
\end{equation}
while the peculiar velocity of the solar system towards the Virgo cluster is
estimated to be\cite{sciama1},\cite{sciama2},\cite{sciama3},%
\begin{equation}
v_{1}=415\text{ }km/s,\text{ \ \ \ }l=335%
{{}^\circ}%
,\text{ \ \ \ }b=7%
{{}^\circ}
\label{sciamaeq1}%
\end{equation}
or%
\begin{equation}
v_{1}=630\text{ }km/s,\text{ \ \ \ }l=330%
{{}^\circ}%
,\text{ \ \ \ }b=45%
{{}^\circ}%
. \label{sciamaeq2}%
\end{equation}
The location of the Virgo cluster is%
\begin{equation}
v=1050\pm200\text{ }km/s,\text{ \ \ \ }l=287%
{{}^\circ}%
,\text{ \ \ \ }b=72.3%
{{}^\circ}%
.
\end{equation}
The infall velocity of the Virgo cluster towards the GA is given as\cite{ga}%
\begin{equation}
v_{in}=1000\pm200\text{ }km/s
\end{equation}
and the GA is located at%
\begin{equation}
v(GA)=4200\text{ }km/s,\text{ \ \ \ }l=309%
{{}^\circ}%
,\text{ \ \ \ }b=18%
{{}^\circ}
\label{gaeq1}%
\end{equation}
or%
\begin{equation}
v(GA)=3000\text{ }km/s,\text{ \ \ \ }l=305%
{{}^\circ}%
,\text{ \ \ \ }b=18%
{{}^\circ}%
. \label{gaeq2}%
\end{equation}
The peculiar velocity of the Virgo cluster relative to the GA is expressed as%
\begin{equation}
v_{2}=1000\pm200\text{ }km/s,\text{ \ \ }l=310.9\pm0.4%
{{}^\circ}%
,\text{ \ \ }b=4.6\pm2.8%
{{}^\circ}%
\text{ \ \ } \label{pv21}%
\end{equation}
for Eq. (\ref{gaeq1}), or%
\begin{equation}
v_{2}=1000\pm200\text{ }km/s,\text{ \ \ }l=307.2\pm1.6%
{{}^\circ}%
,\text{ \ \ }b=1.6\pm3.0%
{{}^\circ}
\label{pv22}%
\end{equation}
for Eq. (\ref{gaeq2}). \ 

With this information, the position of the center, $v_{c}$, can be calculated
as%
\begin{equation}
v(GA)-v_{c}+v_{1}+v_{2}=v(dipole)
\end{equation}
and hence%
\begin{equation}
v_{c}=v(GA)+v_{1}+v_{2}-v(dipole). \label{center}%
\end{equation}
The author estimated the location of the center of the universe to be%
\begin{equation}
v_{c}=5012.2\pm205\text{ }km/s,\text{ \ \ \ }l_{c}=315.3\pm0.2%
{{}^\circ}%
,\text{ \ \ \ }b_{c}=1.7\pm0.3%
{{}^\circ}%
\end{equation}
or%
\begin{equation}
v_{c}=5082.5\pm208\text{ }km/s,\text{ \ \ \ }l_{c}=315.0\pm0.2%
{{}^\circ}%
,\text{ \ \ \ }b_{c}=6.2\pm0.3%
{{}^\circ}%
\end{equation}
for the case of Eq. (\ref{gaeq1}) for the GA, where the two cases correspond
to the velocities of the solar system, Eq. (\ref{sciamaeq1}) or Eq.
(\ref{sciamaeq2}). For the case of Eq. (\ref{gaeq2}) for the GA, one gets
\begin{equation}
v_{c}=4097.6\pm200\text{ }km/s,\text{ \ \ \ }l_{c}=310.7\pm0.2%
{{}^\circ}%
,\text{ \ \ \ }b_{c}=9.4\pm0.4%
{{}^\circ}%
\end{equation}
or%
\begin{equation}
v_{c}=4224.1\pm194\text{ }km/s,\text{ \ \ \ }l_{c}=310.4\pm0.2%
{{}^\circ}%
,\text{ \ \ \ }b_{c}=14.6\pm0.6%
{{}^\circ}
\label{center4}%
\end{equation}
for the two solar system velocities. There are errors in the numerical
estimate in the previous reference\cite{center}, they should be corrected by
the numbers in this article. One notices that the direction obtained for the
center is close to that of Centaurus A,%
\begin{equation}
l(CentaurusA)=309.51587%
{{}^\circ}%
,\text{ \ \ }b(CentaurusA)=19.41732%
{{}^\circ}%
.[.]
\end{equation}
Compare this to the direction in the solution above, Eq. (\ref{center4}).
These results for the location of the center listed above are the corrections
to those in the reference\cite{center}. The above listed velocities correspond
to the distances, 74, 75, 60 and 62 in the units of Mpc. Compare these sizes
with the size of the universe, 14 Gpc.

\section{Apparent center in a B-type universe}

After a general discussion on the universe of B- and C-type, I will come to a
possibility of an apparent center in a B-type universe. As we discussed
earlier, the absence of the center in a B-type universe is definite, since
there is no specific point on a balloon. However, the size of the universe has
increased after the expansion. If the size is small, the gravitational effects
of one portion can be reached to other portion from any direction, in
particular, from the opposite directions, in due time. If the size is
sufficiently large, so that its gravitational effect can be reached only one
direction but not the other direction in the time scale of the age of the
universe, then only a limited amount of portions in the whole universe can
produce a gravitational effect in this time scale. In this case, a finite
amount of the universe, but not the whole universe can give a gravitational
effect. As a result, a finite amount of the universe behaves like a universe.
The center of this finite portion behaves like a center. The only difference
from the center in a C-type universe is that the center defined by the
observer is dependent upon the location of the observer. There is no unique
center for the whole universe. In order to differentiate the center in a
C-type universe and an apparent center of a B-type universe, one has to know
what is the center determined by the other observers. If the center determined
in the previous section is located at a very large distance, one can conclude
that our universe is C-type. Since the distance to the center determined is
relatively small compared to the size of the universe, one would not conclude
whether our universe is C-type or B-type.

\section{Summary and discussion}

The observation of the cmb dipole implies the existence of a center of the
universe. The author has calculated the location of the center in the C-type
Universe and has established that there is an apparent center in a B-type
Universe. In order to determune which Universe we are in, we will need to
obtain data from a different part of the Universe capturing cmb dipole,
peculiar velocity and compute the location of the center. In a C-type
Universe, the author has shown that there should be an unique and universal
position of the center, while the location of the center in a B-type Universe
is dependent on the location of the observer.

\begin{acknowledgments}
The author would like to thank the members of the Physics Department and the
Astronomy Department of the University of Michigan for useful information and
David N. Williams and Peter K. Tomozawa for reading the manuscript.
\end{acknowledgments}

Correspondence should be addressed to the author at tomozawa@umich.edu.


\bigskip\bigskip

\begin{thebibliography}{9}                                                                                                %


\bibitem {dipole1}Melchiorri, B. et. al., New Astron. Rev. 46 (2002) 693.

\bibitem {dipole2}Fixsen, D.J. et. al., Apj. 473 (1996) 576.

\bibitem {center}Tomozawa, Y., MPLA 22 (2007) 1553.

\bibitem {sciama1}Sciama, D. W., Phys. Rev. Letters 18 (1967) 1065.

\bibitem {sciama2}Rees, M. and Sciama, D. W., Nature 213 (1967) 374.

\bibitem {sciama3}Stewart, J. M. and Sciama, D. W., Nature 216 (1967) 748.

\bibitem {ga}Faber, S. M. and Burstein, D., in Large-Scale Motions in the
Universe (ed. Rubin, V. C. and Coyne, G. V., Princeton University Press 1988)
p.115; Lynden-Bell, D. et.al., Apj. 326 (1988) 19.
\end{thebibliography}
\end{document}